\documentclass[pre,twocolumn,superscriptaddress,floatfix,
showpacs,preprintnumbers]{revtex4}

\usepackage[latin2]{inputenc}
\usepackage{amsmath}
\usepackage{amssymb}
\usepackage{epsfig}
\usepackage{bm}
\usepackage{color}

\begin{document}

\title{Evolution of the force distributions in jammed packings of soft particles}

\author{Jens Boberski}
\email{jens.boberski@uni-due.de}
\affiliation{Computational Physics Group, University of
Duisburg-Essen, D-47048 Duisburg, Germany}

\author{M.\ Reza Shaebani}
\email{shaebani@lusi.uni-sb.de}
\affiliation{Department of Theoretical Physics, Saarland 
University, D-66041 Saarbr\"ucken, Germany}

\author{Dietrich E.\ Wolf}
\affiliation{Computational Physics Group, University of
Duisburg-Essen, D-47048 Duisburg, Germany}

\date{\today}

\begin{abstract}
The evolution of the force distributions during the isotropic 
compression of two dimensional packings of soft frictional 
particles is investigated numerically. Regardless of the 
applied deformation, the normal contact force distribution 
$P(f_n)$ can be fitted by the product of a power-law, and a 
stretched exponential, while the tangential force distribution 
$P(f_t)$ is fitted well by a Gaussian. With increasing strain, 
the asymptotic behavior at large forces does not change, but 
both $P(f_n)$ and $P(f_t)$ exhibit a broadening, even though, 
when scaled with the average forces, their widths decrease. 
Furthermore, the distribution of friction mobilization 
$P(\eta)$ is a decreasing function of $\eta=|f_t|/(\mu f_n)$, 
except for an increased probability of fully mobilized 
contacts ($\eta{=}1$). The excess coordination number of the 
packings increases with the applied strain, indicating that 
the more a packing is compressed the more stable it becomes.
\end{abstract}


\maketitle
\section{Introduction}
When subject to external forces, disordered materials 
such as emulsions, colloidal suspensions, and granular 
media exhibit a nontrivial response, which has received 
considerable attention over the last two decades 
\cite{Neddermann92,Jaeger96}. Highly heterogeneous 
force networks form between the particles \cite{Snoeijer04,
Tighe05} and one of the main concerns regarding the 
contact force statistics is their asymptotic behavior 
at large forces.

Early experimental measurements \cite{Liu95,ExpTailRefs,
Erikson02} and numerical simulations with contact dynamics 
\cite{Radjai96} or soft particle methods with Hookian 
\cite{Landry03,Silbert02} and Hertzian \cite{Liu95,
Silbert02} interactions reported an exponential tail 
for the normal force distribution $P(f_n)$. However, 
faster than exponential tails were also found later in 
experiments \cite{Makse00,Majmudar05,Zhou06} and 
simulations with Hookian \cite{Ohern02}, Hertzian 
\cite{Makse00,Ohern02,Zhang05} and other force laws 
\cite{Ohern01,Brujic03}. There have been theoretical 
attempts to explain the tail behavior \cite{TheoreticalRefs} 
which mainly support the exponential decay. More recently, 
Tighe et al.\ \cite{Tighe2} took the force balance at 
the particle level into account when maximizing the 
entropy with respect to the admissible force states 
and proposed an analytic expression for the normal 
force distribution in static granular media, 
predicting a Gaussian behavior for large forces, 
supported by a numerical study \cite{vanEerd07} 
within the framework of the force network ensemble 
\cite{Snoeijer04,Ensemble1}. Some of the studies 
observe a crossover form an exponential-like to a 
faster than exponential behavior with increasing 
the applied isotropic compression \cite{Ohern02,
Makse00,Nguyen00,Howell99,Zhang05, Hidalgo09}. In 
this paper we provide numerical evidence for the 
faster than exponential decay in frictional soft 
spheres, and address the question whether the 
asymptotic behavior changes qualitatively when 
approaching the jamming transition.

Much less has been reported so far on the tangential 
force distribution $P(f_t)$ \cite{Majmudar05,Hidalgo09,
Radjai96,Silbert02,Zhang10} and the evolution of the 
force distributions during deformation processes. Here, 
we study the evolution of the tangential force 
distribution as well as the friction mobilization 
during the isotropic compression. Note that for example the 
experiments on photoelastic particles provide evidence for a different behavior 
of the force distributions in sheared systems. For 
example, $P(f_n)$ of sheared packings decays much 
more slowly than that of the compressed ones. Besides 
the path dependence of the deformation, shear 
induced anisotropy has been also reported \cite{Majmudar05,
Zhang10}. The study of sheared systems is however beyond the 
scope of this work.

\section{Simulation method}
In this report, the evolution of the inter-particle forces during 
a quasi-static isotropic compression was studied numerically (using 
DEM simulations). The two dimensional simulation box has periodic
boundaries to avoid effects due to the presence of walls. The 
particle interactions are computed, using a linear 
spring-dashpot model for both normal and tangential forces (with 
the spring constant ratio $k_t/k_n$ either being $0.5$ or $1$), 
as well as the Hertz-Mindlin model as described in \cite{hertz}. 
The tangential force is additionally limited by the Coulomb limit. 
The results presented throughout the paper belong to the linear 
force law with friction coefficient $\mu{=}0.5$ and stiffness 
ratio $k_t/k_n{=}0.5$, unless stated otherwise.

The packings consist of nearly $14,200$ disks with radii taken 
from a uniform distribution in the range $[0.8 \bar r,1.2 \bar r]$, 
where $\bar r$ is the average particle radius and is used as the 
natural unit of length in the following. The unit of force is 
chosen to be $k_n \bar r$.

The initial configuration is generated by randomly placing the 
particles without accepting any overlap between them. Afterwards, 
this unjammed system is compressed quasi-statically through a 
sequence of incremental compression and relaxation steps. A 
compression step is realized by re-scaling the particle 
positions and keeping their radii fixed. The relaxation routine 
ensures that the net force exerted on each particle is a factor 
of $10^{-8}$ below the mean contact force, before the next 
deformation step starts. Each time the system is equilibrated, 
the force state of the packing is stored. This process continues 
until the average overlap exceeds a given value. Upon decreasing 
the volume, the average overlap $\langle \xi_n \rangle$ remains 
zero until the jamming transition is reached. Beyond this 
transition $\langle \xi_n \rangle$ increases with the compression 
and, due to its purely geometric origin, is chosen as the order 
parameter to characterize the jammed state. Note that O'Hern et 
al. \cite{Ohern02} suggested that averaging over configurations with slightly 
different average forces can distort the results close to the 
jamming transition. To avoid such effects, the 
calculated distributions in our study are taken from single 
realizations.

\section{Numerical Results}
\subsection{Normal force distribution}
\begin{figure}[t]
\centering
\includegraphics[width=0.9\columnwidth]{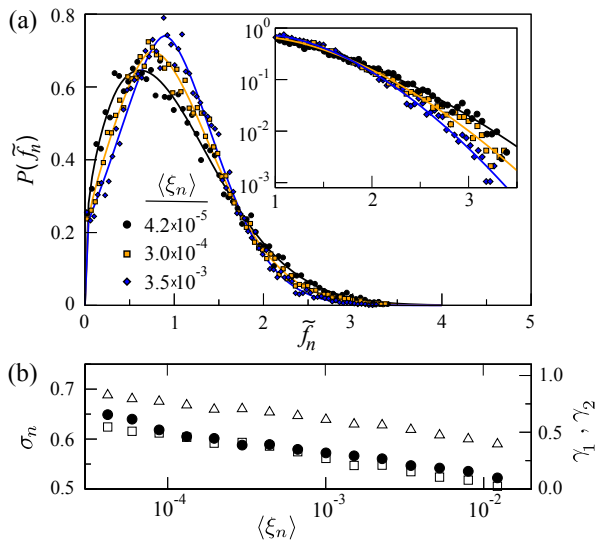}
\caption{(Color online) (a) Distribution of the normalized normal 
forces $\tilde{f}_n {=} f_n/\langle f_n\rangle$ for increasing 
applied deformation. Inset: the same plot in log-lin scale. The 
lines are fits using Eq.~(\ref{Eq:PnForm}). (b) The standard 
deviation $\sigma_n$ (full circles), the skewness $\gamma_{_1}$ 
(open triangles), and the excess kurtosis $\gamma_{_2}$ (open 
squares) versus $\langle \xi_n \rangle$.}
\label{Figure:1}
\end{figure}
Upon increasing the deformation in our simulations, not only the 
mean value of $f_n$ increases but also the shape of the distribution 
is affected, e.g., the standard deviation and skewness change. Using 
the normalized contact force $\tilde{f_n} {\equiv} f_n / \langle f_n 
\rangle$, three typical distributions $P(\tilde{f}_n)$ at different 
values of $\langle \xi_n \rangle$ are shown in Fig.~\ref{Figure:1}(a). 
In all cases, the distribution has a peak and one can see that the 
tail of the distributions gradually becomes more bent during the 
compression process. These results, as well as the observations for 
the tangential forces and mobilization (discussed below) are in good 
agreement with experimental results by Majmudar and Behringer 
\cite{Majmudar05}. Fits of the form 
\begin{equation}
P(\tilde f_n) =\frac{\text{1}}{N} \, 
({\tilde f_n})^{\nu_n} \,\exp\left(-
\left|\frac{\tilde f_n - b_n}{w_n} \right|^{\delta_n}\right)\,,
\label{Eq:PnForm}
\end{equation} 
weighted with the reciprocal variance, are found to characterize the 
shape of $P(\tilde f_n)$ for different compressions (see lines in 
Fig.~\ref{Figure:1}(a)). The fitting parameters $N$, $\nu_n$, $w_n$, 
$b_n$, and $\delta_n$ are not independent, due to constrains for the 
normalization $\int_0^\infty P(\tilde f_n)\text{d}\tilde f_n{=}1$ and 
on the first moment $\int_0^\infty \tilde f_n P(\tilde f_n)\text{d} 
\tilde f_n{=}1$ of the distribution. Thus the number of fit parameters 
can be reduced to three. Even without taking the constrains into 
account during fitting, they are fulfilled with deviations below 
$0.5\%$.

\begin{figure}[t]
\centering
\includegraphics[width=0.9\columnwidth]{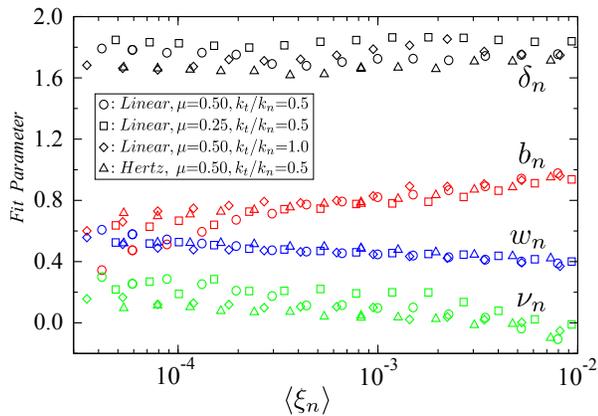}
\caption{(Color online) The compression dependence of the fitting 
parameters $\delta_n$, $b_n$, $w_n$, and $\nu_n$ according to 
Eq. \eqref{Eq:PnForm} for systems with different contact force  
properties.}
\label{Figure:2}
\end{figure}

Note that the fit in Eq.~(\ref{Eq:PnForm}) captures not only the 
tail behavior but also the overall shape of the distribution with 
effectively three fitting parameters. The robustness of the results 
with respect to the microscopic properties of the interparticle 
force is studied, using different values of friction coefficient 
$\mu$ and stiffness ratio $k_t/k_n$, as well as linear and 
non-linear force laws. The results for four sets of the parameters 
are shown in Fig~\ref{Figure:2}. While the exact values of the 
fitting parameters depend on the contact force properties, the 
qualitative behavior in terms of the mean normal overlap is 
universal. For example, the exponent $\delta_n$ is found to be 
independent of the compression with values ranging from $1.65$ to 
$1.8$, depending on the contact properties. Allowing for this 
additional parameter leads to fits that capture the tail behavior 
substantially better than Gaussian ($\delta_n{=}2$) or exponential 
($\delta_n{=}1$) fits. This supports the reports on a decay faster 
than exponential for large forces and shows that while the shape 
of the distribution changes as the jamming transition is approached, 
the asymptotic behavior for large forces remains the same.

The evolution of the width $w_n$ with increasing deformation is 
shown in Fig.~\ref{Figure:2}, indicating that the normalized 
distributions become narrower. The figure also shows that the 
exponent $\nu_n$ decreases with $\langle \xi_n \rangle$, while 
the shift $b_n$ increases with the compression. In 
Fig.~\ref{Figure:1}(b), the shape change of $P(\tilde f_n)$ is 
investigated by calculating higher moments of the force sets. 
The standard deviation, $\sigma_n$, of the data decreases with 
$\langle \xi_n \rangle$, similar to the behavior of $w_n$. A 
smaller standard deviation of the normalized forces at higher 
compression corresponds to a more homogeneous force network. 
For comparison, let us consider a situation where all contact 
forces are just rescaled, when increasing the external load. 
In this case, the relative width of the force distribution 
would remain constant. The fact that $\sigma_n$ decreases may 
be attributed to the possibility of opening and closing of 
new contacts during the compression, and also to the nonaffine 
motions which allow local changes of the overlaps leading to 
a more uniform stress at larger deformation.

The skewness $\gamma_{_1}$, which reflects the degree of 
asymmetry of a distribution (for $\gamma_{_1}{=}0$ being 
symmetric) is shown in Fig.~\ref{Figure:1}. It is a decreasing 
function of $\langle \xi_n \rangle$, thus, the distributions 
become more symmetric at larger deformations. The excess kurtosis 
$\gamma_{_2}$ describes the peakedness and tail behavior of 
the distribution. A higher $\gamma_{_2}$ represents a fatter 
tail (slower decay). Therefore, the decrease of $\gamma_{_2}$ 
in Fig.~\ref{Figure:1}(b) agrees with the observation that 
the bending of the tail increases, when the packing is more 
compressed. The behavior for $\gamma_1$ and $\gamma_2$ 
may have led to the impression of a crossover between an exponential 
and Gaussian tail in previous publications.

\subsection{Tangential force distribution}
\begin{figure}[t]
\centering
\includegraphics[width=0.9\columnwidth]{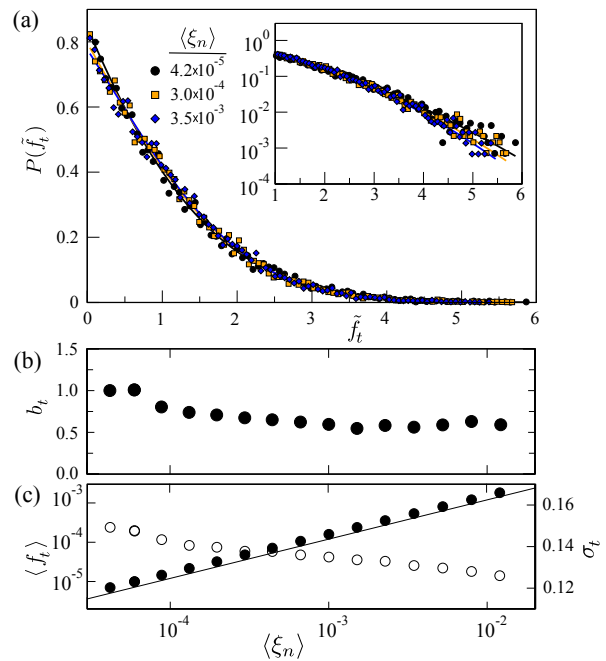}
\caption{(Color online) (a) Distribution of normalized 
tangential forces $P(\tilde{f}_t)$ for different values 
of the mean overlap $\langle \xi_n \rangle$. Inset: same 
in log-lin scale. The lines indicate fits given by 
Eq.~(\ref{Eq:PtForm}). (b) The shift $b_t$ of the 
Gaussian versus $\langle \xi_n \rangle$. (c) The average 
tangential force $\langle f_t \rangle$ (full circles) 
and the standard deviation $\sigma_t$ (open circles) versus 
$\langle \xi_n \rangle$. The solid line corresponds to a 
linear increase.}
\label{Figure:3}
\end{figure}
The distribution of tangential forces is a monotonically 
decreasing function of $f_t$, without a notable peak or 
plateau. Figure \ref{Figure:3}(a) shows that the data nearly 
collapse for different compressions, if the forces are 
normalized with the mean force $\tilde{f_t} {\equiv} |f_t| 
/\langle |f_t| \rangle$, which means that the distribution 
broadens with increasing strain. The semi-logarithmic inset 
shows an increasing curvature with increasing average 
overlap (even though much less pronounced than for$P(\tilde 
f_n)$), indicating a non-exponential behavior. The data 
over the whole range of tangential forces can be fitted with
 \begin{equation}
P({\tilde f}_t)  = \displaystyle\frac{\text{1}}{N} \, 
\text{exp}\left[-\left(\frac{\tilde f_t}{ 
w_t} - b_t\right)^\text{2}\right], 
\label{Eq:PtForm}
\end{equation}
where $b_t$ is the fitting parameter and the normalization 
gives $N{=}\frac 12 w_t \sqrt{\pi} (1 + \text{erf}(b_t))$, 
while the constraint for the first moment leads to 
\begin{equation}
w_t=\frac{1 + \text{erf}(b_t)}{b_t+b_t\text{erf}(b_t) + 
\frac{1}{\sqrt{\pi}} \exp(-2b_t)}\,. 
\end{equation}
Adding a variable exponent (like $\delta_n$ in \eqref{Eq:PnForm}) 
in the exponential function as an additional parameter does not 
improve the fit. Adding an additional power law term, leads to a 
small, negative exponent (${\sim}-0.05$) and thus an almost 
constant pre-factor. An alternative fit to an exponential which 
has been proposed in the literature is only applicable for the 
tail in the low deformation regime and can not be supported by 
the data in this study. 

Figure \ref{Figure:3}(b) shows that $b_t$ first decreases with 
compression and then reaches an overlap-independent regime 
for overlaps larger than $10^{-3}$. The standard deviation 
$\sigma_t$ of the normalized tangential forces, shown in figure
\ref{Figure:3}(c), decreases by about $25\%$ similarly to $\sigma_n$. 
The average tangential force increases linearly with increasing 
overlap (see Fig.~\ref{Figure:3}(c)).

\begin{figure}[t]
\centering
\includegraphics[width=0.99\columnwidth]{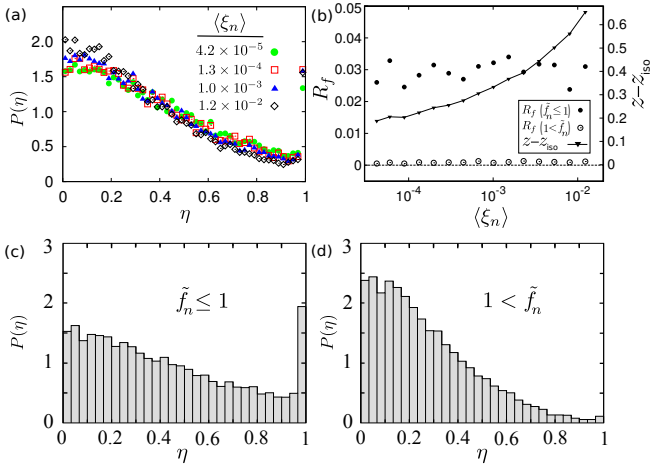}
\caption{(Color online) (a) Distribution of friction 
mobilization $P(\eta)$ for different strains. (b) 
Fraction of the nearly mobilized contacts $R_f$ 
(separately shown for small and large normal forces), 
and the distance from isostaticity $z{-}z_\text{iso}$ 
in terms of $\langle \xi_n \rangle$ for the linear 
force law with $\mu{=}k_t/k_n{=}0.5$. (c),(d) Histograms 
of the friction mobilization $\eta$ for small and 
large forces in the packing with $\langle \xi_n 
\rangle {=} 1.2 \times 10^{-2}$.} 
\label{Figure:4}
\end{figure}

\subsection{ Friction mobilization} 
To elucidate the influence of friction on the evolution of the 
contact forces, the distribution of friction mobilization $P(\eta)$ 
is studied. The Coulomb condition specifies that $|f_t| \leq \mu 
f_n$. Therefore, the mobilization $\eta{=}|f_t| {/} \mu f_n$ 
quantifies the distance of a contact from the Coulomb friction 
limit. It varies within the range $[0,1]$, with $\eta {=} 1$ 
for fully mobilized contacts, i.e.\ those contacts in the static 
packing that are on the verge of sliding. Figure \ref{Figure:4}(a) 
shows that $P(\eta)$ decays with increasing $\eta$ except for a 
peak at $\eta {\approx} 1$. Similar results were observed in 
experiments with photoelastic particles \cite{Majmudar05}. The 
decay of $P(\eta)$ shows that most of the contacts are far from 
the Coulomb limit. With increasing deformation, $P(\eta)$ grows 
for small and decreases for large $\eta$, i.e.\ the probability 
distribution becomes steeper. The probability of fully mobilized 
contacts, however, remains approximately independent of $\langle 
\xi_n \rangle$. This is shown more clearly in Fig.~\ref{Figure:4}(b), 
where the fraction $R_f$ of contacts with a mobilization close 
to one ($\eta {>} 0.99$) is separately calculated for weak 
($\tilde{f}_n {\leq} 1$) and strong ($1 {<} \tilde{f}_n$) normal 
forces, since the underlying mechanisms of stress propagation 
have been found to be different for these two subnetworks 
\cite{WeakStrongNetworks}. $R_f$ fluctuates around $3\%$ ($0.1\%$) 
for weak (strong) forces. Note that these values depend in general 
on the friction coefficient \cite{Shundyak07}. The constant nature 
of $R_f$ together with the increase of the average coordination 
number $z$ with $\langle \xi_n \rangle$ indicates that the excess 
coordination number of the packing $\Delta z {=}z{-} z_\text{iso}$ 
grows when increasing the applied deformation [Fig.~\ref{Figure:4}(b)]. 
It was shown that the mechanical response of frictional packings 
exhibits a critical scaling with $\Delta z$ in the limit of $\Delta 
z{\rightarrow}0$ \cite{Somfai07}. More generally, $\Delta z$ 
influences the extent of force indeterminacy which governs 
the stability of the granular packings \cite{Shaebani07}. 
The enhancement of mechanical stability with increasing the 
 compression can be also traced back to the fact that 
topological properties of the force network, like the number 
of triangular structures, evolve during the process 
\cite{Arevalo10}, allowing to predict the jamming transition 
point. The triangular structures are found to play an important 
role in stabilizing the packing \cite{Tordesillas10}.

The majority of the contacts with a mobilization close to $1$ 
carry a normal force below the average. Figures~\ref{Figure:4}(c) 
and (d) show the distribution of $\eta$ separately for both 
categories of forces. In addition to the difference at 
$\eta{\approx}1$, the histogram is steeper for strong forces. 
This shows that the force network of the strong forces is more 
stable and sliding events mainly occur at weak contacts. 

\begin{figure}[t]
\centering
\includegraphics[width=0.99\columnwidth]{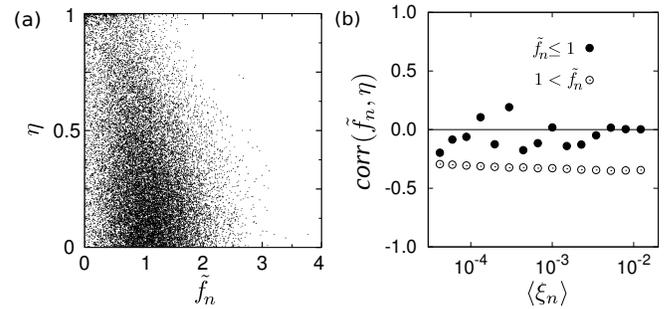}
\caption{(Color online) (a) A scatter plot of $\eta$ 
versus $\tilde f_n$ for the packing with $\langle \xi_n 
\rangle {=} 1.9 \times 10^{-3}$. (b) The correlation between 
$\tilde f_n$ and $\eta$ as a function of $\langle \xi_n 
\rangle$, separately shown for small (full circles) and 
large (open circles) forces.} 
\label{Figure:5}
\end{figure}

The simulation results reveal a dependence of the friction 
mobilization at a contact on the magnitude of the normal 
force carried by that contact. Figure~\ref{Figure:5}(a) shows 
a typical scatter plot of $\eta$ versus $\tilde{f}_n$. The 
accumulation of contacts with $\eta =1$ denotes the fully 
mobilized contacts at the edge of the Coulomb friction cone. 
For weak normal forces, the data points are scattered over 
the whole range of possible mobilizations. However, at 
strong normal forces, they tend towards smaller values of 
$\eta$ and the fraction of fully mobilized contacts diminishes. 
This results in the decrease of the average mobilization 
as a function of the normal force (not shown). Thus the 
contacts which carry small normal forces are more liable 
for non-elastic deformations not only due to the higher 
chance of opening for such contacts (because of a smaller 
overlap), but also due to sliding, since these contacts 
have a larger mobilization. To quantify the relation, 
the Pearson correlation coefficient of $f_n$ and $\eta$ 
is shown in Fig.~\ref{Figure:5}(b). The correlation is 
calculated for each packing, separately for weak and 
strong forces. While the correlations are scattered 
around zero for weak normal forces, there is a robust 
anti-correlation between $f_n$ and $\eta$ for strong 
forces, which remains unchanged over the whole range 
of deformations. This verifies again that $f_n$ and 
$\eta$ are inversely related to each other, and that 
larger forces show a lower probability to exceed the 
sliding threshold than small forces.

\section{Conclusions}
The inter-particle force evolution during the isotropic 
compression of soft particle packings was investigated 
and empirical expressions for the normal and tangential 
force distributions were presented, that showed a faster 
than exponential decay for large forces. The results 
revealed that the relative width of the distributions
decrease with increasing deformation, leading to a 
more homogeneous force network. It is notable that while 
the width of the distribution changes, the decay of the 
large forces remains unchanged. The investigation of 
friction mobilization $\eta$ showed that, independent 
of the deformation, the fully mobilized contacts are 
most likely those carrying small normal forces. Finally, 
we note that the investigation of sheared systems is 
essential as a further step towards understanding the 
deformation at the microscopic level, which complements 
the theoretical studies based on the assumption of 
affine motion of particles \cite{Boberski12} and 
facilitates the development of a general microscopic 
model for the deformation of granular materials.

\begin{acknowledgments}
We would like to thank Lothar Brendel for helpful 
discussions and carefully reading our manuscript. We 
would like to acknowledge the support by the German 
Research Foundation (DFG) via priority program SPP 1486 
"Particles in Contact" and the Center for Computational 
Sciences and Simulation of the University of Duisburg-Essen.
\end{acknowledgments}


\begin{thebibliography}{10}
\bibitem{Neddermann92} R. M. Neddermann, {\it Statics and 
Kinematics of Granular Materials} (Cambridge University Press, 
Cambridge, 1992).

\bibitem{Jaeger96} H. M. Jaeger, S. R. Nagel, and R. P. Behringer, 
Rev. Mod. Phys. \textbf{68}, 1259 (1996).

\bibitem{Snoeijer04} J. H. Snoeijer, T. J. H. Vlugt, M. 
van Hecke, and W. van Saarloos, Phys. Rev. Lett. \textbf{92}, 
054302 (2004).

\bibitem{Tighe05} B. P. Tighe, J. E. S. Socolar, D. G. Schaeffer, 
W. G. Mitchener, and M. L. Huber, Phys. Rev. E \textbf{72}, 031306 
(2005).

\bibitem{Liu95} C. Liu, S. R. Nagel, D. A. Schecter, S. N. 
Coppersmith, S. Majumdar, O. Narayan, and T. A. Witten, Science 
\textbf{269}, 513 (1995).

\bibitem{Erikson02} J. M. Erikson, N. W. Mueggenburg, H. M. Jaeger, 
and S. R. Nagel, Phys. Rev. E \textbf{66}, 040301(R) (2002).

\bibitem{ExpTailRefs} D. L. Blair, N. W. Mueggenburg, A. H. Marshall, 
H. M. Jaeger, S. R. Nagel, Phys. Rev. E \textbf{63}, 041304 (2001); 
G. L\o{}voll, K. J. M\aa{}l\o{}y, and E. G. Flekk\o{}y, Phys. Rev. E 
\textbf{60}, 5872 (1999); D. M. Mueth, H. M. Jaeger, and S. R. Nagel, 
Phys. Rev. E \textbf{57}, 3164 (1998).

\bibitem{Radjai96} F. Radjai, M. Jean, J. J. Moreau, S. Roux, Phys. 
Rev. Lett. \textbf{77}, 274 (1996).

\bibitem{Landry03} J. W. Landry, G. S. Grest, L. E. Silbert, and S. 
J. Plimpton, Phys. Rev. E \textbf{67}, 041303 (2003).

\bibitem{Silbert02} L. E. Silbert, G. S. Grest, and J. W. Landry, 
Phys. Rev. E \textbf{66}, 061303 (2002).

\bibitem{Majmudar05} T. S. Majmudar and R. P. Behringer, Nature 
\textbf{435}, 1079 (2005).

\bibitem{Makse00} H. A. Makse, D. L. Johnson, and L. M. Schwartz, 
Phys. Rev. Lett. \textbf{84}, 4160 (2000).

\bibitem{Zhou06} J. Zhou, S. Long, Q. Wang, and A. D. Dinsmore, 
Science \textbf{312}, 1631 (2006).

\bibitem{Ohern02} C. S. O'Hern, S. A. Langer, A. J. Liu, and S. R. 
Nagel, Phys. Rev. Lett.  \textbf{88}, 075507 (2002).

\bibitem{Zhang05} H. P. Zhang and H. A. Makse, Phys. Rev. E \textbf{72}, 
011301 (2005).

\bibitem{Ohern01} C. S. O'Hern, S. A. Langer, A. J. Liu, and S. R. 
Nagel, Phys. Rev. Lett. \textbf{86}, 111 (2001)

\bibitem{Brujic03} J. Brujic, S. F. Edwards, I. Hopkinson, and 
H. A. Makse, Physica A \textbf{327}, 201 (2003); 

\bibitem{TheoreticalRefs} S. N. Coppersmith, C.-h. Liu, S. Majumdar, 
O. Narayan, T. A. Witten, Phys. Rev. E \textbf{53}, 4673 (1996); 
J. Rottler and M. O. Robbins, Phys. Rev. Lett. \textbf{89}, 195501 
(2002); P. T. Metzger, Phys. Rev. E \textbf{70}, 051303 (2004); 
K. Bagi, Granular Matter \textbf{5}, 45 (2003); N. P. Kruyt and 
L. Rothenburg, Int. J. Solids Struct. \textbf{39}, 571 (2002).

\bibitem{Tighe2} B. P. Tighe, A. R. T. van Eerd, and T. J. H. Vlugt, 
Phys. Rev. Lett. \textbf{100}, 238001 (2008).

\bibitem{vanEerd07} A. R. T. vanEerd, W. G. Ellenbroek, M. vanHecke, 
J. H. Snoeijer, and T. J. H. Vlugt, Phys. Rev. E \textbf{75}, 
060302(R) (2007).

\bibitem{Ensemble1} J. N. Roux, Phys. Rev. E \textbf{61}, 6802 (2000); 
L. E. Silbert, D. Ertas, G. S. Grest, T. C. Halsey, D. Levine, Phys. 
Rev. E \textbf{65}, 031304 (2002); T. Unger, J. Kertesz, and D. E. 
Wolf, Phys. Rev. Lett. \textbf{94}, 178001 (2005); M. R. Shaebani, 
T. Unger, and J. Kertesz, Phys. Rev. E \textbf{79}, 052302 (2009).

\bibitem{Hidalgo09} R. C. Hidalgo, I. Zuriguel, D. Maza, I. 
Pagonabarraga, Phys. Rev. Lett. \textbf{103}, 118001 (2009).

\bibitem{Howell99} D. Howell, R. P. Behringer, and C. Veje, Phys. 
Rev. Lett. \textbf{82}, 5241 (1999).

\bibitem{Nguyen00} M. L. Nguyen and S. N. Coppersmith, Phys. Rev. E 
\textbf{62}, 5248 (2000).

\bibitem{Zhang10} J. Zhang, T. S. Majmudar, M. Sperl and R. P. 
Behringer, Soft Matter \textbf{6}, 2982 (2010).

\bibitem{hertz} L. E. Silbert, D. Ertas, G. S. Grest, T. C. Halsey, 
D. Levine, S. J. Plimpton, Phys. Rev. E \textbf{64}, 051302 (2001).

\bibitem{WeakStrongNetworks} F. Radjai, D. E. Wolf, M. Jean, and 
J.-J. Moreau, Phys. Rev. Lett. \textbf{80}, 61 (1998).

\bibitem{Shundyak07} K. Shundyak, M. van Hecke, and W. van Saarloos, 
Phys. Rev. E \textbf{75}, 010301(R) (2007).

\bibitem{Somfai07} E. Somfai, M. van Hecke, W. G. Ellenbroek, K. 
Shundyak, and W. van Saarloos, Phys. Rev. E \textbf{75}, 020301(R) 
(2007).

\bibitem{Shaebani07} M. R. Shaebani, T. Unger, and J. Kertesz, 
Phys. Rev. E \textbf{76}, 030301(R) (2007); Phys. Rev. E \textbf{78}, 
011308 (2008); J. H. Snoeijer, W. G. Ellenbroek, T. J. H. Vlugt, 
and M. van Hecke, Phys. Rev. Lett. \textbf{96}, 098001 (2006).

\bibitem{Arevalo10} R. Ar\'evalo, I. Zuriguel, and D. Maza, Phys. 
Rev. E \textbf{81}, 041302 (2010).

\bibitem{Tordesillas10} A. Tordesillas, D. M. Walker, and Q. Lin, 
Phys. Rev. E \textbf{81}, 011302 (2010).

\bibitem{Boberski12} M. R. Shaebani, J. Boberski, and D. E. Wolf, 
Granular Matter \textbf{14}, 265 (2012).

\end{thebibliography}
\end{document}